%% file: main.tex
\documentclass[amsmath,amssymb,twocolumn,notitlepage,nofootinbib,superscriptaddress,floatfix,eprint]{revtex4-2}
\usepackage[utf8]{inputenc}
\usepackage[T1]{fontenc} 
\usepackage{graphicx}
\usepackage{braket}
\usepackage{hyperref}
\usepackage{xcolor}
\usepackage{float}
\usepackage{bm}
\usepackage{soul}
\usepackage{mathtools}
\usepackage{microtype}
\usepackage{subcaption} 
\usepackage[english]{babel}

\usepackage{xcolor}
\usepackage{overpic} 
\usepackage{array}[=2016-10-06]

\begin{document}

\title{Compact and Long-Term Stable Polarization Entangled Photon-Pair Source}

\author{Mirela~Selimovi\'{c}}
\email{mirela.selimovic@univie.ac.at}
\affiliation{zerothird GmbH, Clemens-Holzmeister-Straße 6/6, 1100 Vienna, Austria}
\affiliation{University of Vienna, Faculty of Physics, Vienna Center for Quantum
Science and Technology (VCQ), Boltzmanngasse 5, Vienna 1090, Austria}
\affiliation{University of Vienna, Faculty of Physics, Vienna Doctoral School in Physics (VDSP), Boltzmanngasse 5, 1090 Vienna, Austria}

\author{Josip~Bajo}
\affiliation{University of Vienna, Faculty of Physics, Vienna Center for Quantum
Science and Technology (VCQ), Boltzmanngasse 5, Vienna 1090, Austria}
\affiliation{University of Vienna, Faculty of Physics, Vienna Doctoral School in Physics (VDSP), Boltzmanngasse 5, 1090 Vienna, Austria}

\author{Mikl\'{o}s~Horv\'{a}th}
\affiliation{qtlabs GmbH,  Clemens-Holzmeister-Straße 6/6, 1100 Vienna, Austria}

\author{Juan~Jos\'e~Romero}
\affiliation{Universidad Polit\'ecnica de Madrid, Madrid, Spain}
\affiliation{Center for Computational Simulation, Madrid, Spain}

\author{Thomas~Heine}
\affiliation{zerothird GmbH, Clemens-Holzmeister-Straße 6/6, 1100 Vienna, Austria}

\author{Philip~Walther}
\affiliation{University of Vienna, Faculty of Physics, Vienna Center for Quantum
Science and Technology (VCQ), Boltzmanngasse 5, Vienna 1090, Austria}
\affiliation{University of Vienna, Faculty of Physics, Vienna Doctoral School in Physics (VDSP), Boltzmanngasse 5, 1090 Vienna, Austria}

\author{Rupert~Ursin}
\affiliation{zerothird GmbH, Clemens-Holzmeister-Straße 6/6, 1100 Vienna, Austria}
\affiliation{qtlabs GmbH,  Clemens-Holzmeister-Straße 6/6, 1100 Vienna, Austria}

\author{Sebastian~Philipp~Neumann}
\email{sebastian.neumann@zerothird.com}
\affiliation{zerothird GmbH, Clemens-Holzmeister-Straße 6/6, 1100 Vienna, Austria}

\begin{abstract}
We present a robust, compact polarization-entangled photon-pair source based on bulk optics that achieves long-term stability, a spectral pair generation rate of $4.24 \times 10^6$\,pairs s$^{-1}$ mW$^{-1}$ nm$^{-1}$ constant for pump powers up to 700\,mW, an intrinsic heralding efficiency of 59.8\,\% , and a polarization contrast visibility of 99.6\,\%. The source is mechanically designed for vibrational robustness, employing an engineered base plate and a reduced optical layout for less alignment sensitivity and complexity. It operates at telecom wavelengths in C-band with a typical type-0 SPDC spectral output of 70\,nm width. We demonstrate its integration into a deployed entanglement-based QKD system that bridges the gap between laboratory prototypes and practical quantum communication hardware.

\end{abstract}

\maketitle

\section{Introduction}
As quantum technologies mature beyond laboratory demonstrations, there is a need for compact, stable, and manufacturable hardware. Single and entangled photons are essential resources of quantum cryptography, metrology, computation and studying fundamental physics of quantum mechanics \cite{bozzio_quantum_2025, couteau_applications_2023, giustina2015significant, maring2024versatile, psiquantum2025manufacturable, kviatkovsky_microscopy_2020}. 

Photon sources can broadly be divided into probabilistic and non-probabilistic, i.e. on-demand sources. On-demand single-photon sources are based on individual quantum emitters and can, in principle, generate a single photon upon excitation with a well-defined emission cycle. Prominent platforms include semiconductor quantum dots \cite{schimpf_quantum_2021, Loredo2026, ding2025high}, color centers in diamonds such as nitrogen-vacancy (NV) centers \cite{kim_quantum_2025}, silicon-carbide color centers based on spin-photon interfaces \cite{bader2024analysis}, as well as cavity coupled trapped neutral atoms and ions, enabling controlled emission of matter-photon entanglement and multiphoton quantum states \cite{canteri2025photon, thomas2024fusion}. Recently, more emerging platforms have demonstrated deterministic single photons emission including two-dimensional materials \cite{chhaperwal2026origin, gerard2026resonance}, carbon nano-tubes \cite{husel2024cavity}, rare-earth ions for spin-photon entanglement \cite{uysal2025spin}, and silicon color centers \cite{Saggio2024}.

In contrast, probabilistic photon-pair sources rely on nonlinear optical processes, such as spontaneous parametric down-conversion (SPDC) or spontaneous four-wave mixing. Conventional and established bulk platforms include beta-barium borate (BBO), periodically-poled potassium titanyl phosphate (ppKTP), and periodically-poled lithium niobate (ppLN) \cite{kwiat1995new, fedrizzi2007wavelength, finco2024time}, and more mature platforms based on pure silicon \cite{kramnik2025scalable}, silicon nitride \cite{li2025down} and silicon carbide \cite{rahmouni2024entangled}. Recent efforts have demonstrated photon pairs in nonlinear metasurfaces \cite{santiago2022resonant, zhang2022spatially}, novel ultrathin materials \cite{wang2026nonlinear, weissflog2024tunable, ma2026twin, lyu2025tunable} and periodically poled layered semiconductors \cite{trovatello2025quasi, braun2026scaling}. 
Integrated on-chip \cite{shi_integrated_2026} and waveguide-based approaches \cite{alibart_quantum_2016, zhuang_ultrabright_2025, troisi_high-brightness_2026} offer promising advantages in terms of compactness and scalability. Nevertheless, bulk optic sources of entangled photon pairs are still a robust approach for generating high-quality polarization-entangled photon pairs mainly because of superior coupling efficiency. In particular, entanglement-based quantum key distribution (eQKD) protocols such as BBM92 \cite{gisin_quantum_crypto_2002} require reliable sources of entangled photon pairs that can operate outside of laboratory conditions. To meet this demand, photon-pair sources must transition from sensitive optical table setups to reliable and stable modules without compromising performance. In this work, we present a robust, compact entangled photon-pair source based on type-0 SPDC with a quasi phase-matched periodically-poled lithium niobate (QPM ppLN) bulk crystal in the well established bulk optics Sagnac interferometric configuration \cite{anwar_entangled_2021}. The source interferometer is pumped with a continuous-wave (CW) laser producing polarization-entangled photon pairs at telecom wavelengths in C-band. It is designed for mechanical stability, with an engineered base plate and minimal optical layout that promises to reduce alignment drift, photon loss, and footprint. It achieves high polarization visibilities of 99.6\,\%, a spectral pair generation rate of $4.24\times10^6$\,pairs s$^{-1}$mW$^{-1}$nm$^{-1}$, and an intrinsic  heralding of up to 59.8\,\%, while offering a broad emission spectrum of about 70\,nm compatible with existing telecommunication infrastructure for both point-to-point and wavelength-division multiplexed quantum networks \cite{wengerowsky_entanglement-based_2018, valencia_large-scale_2026, shi_integrated_2026}. The long-term stability of the source has been demonstrated by its deployment for 12 months as part of an eQKD system under the realistic conditions of the MadQCI ecosystem \cite{martin_madqci_2024, sebastian-lombrana_scaling_2026}.

\section{Source Design and Characterization}

\begin{figure*}[htb]
    \centering

    \begin{minipage}[c]{0.55\textwidth}
        \centering
        \includegraphics[width=\linewidth]{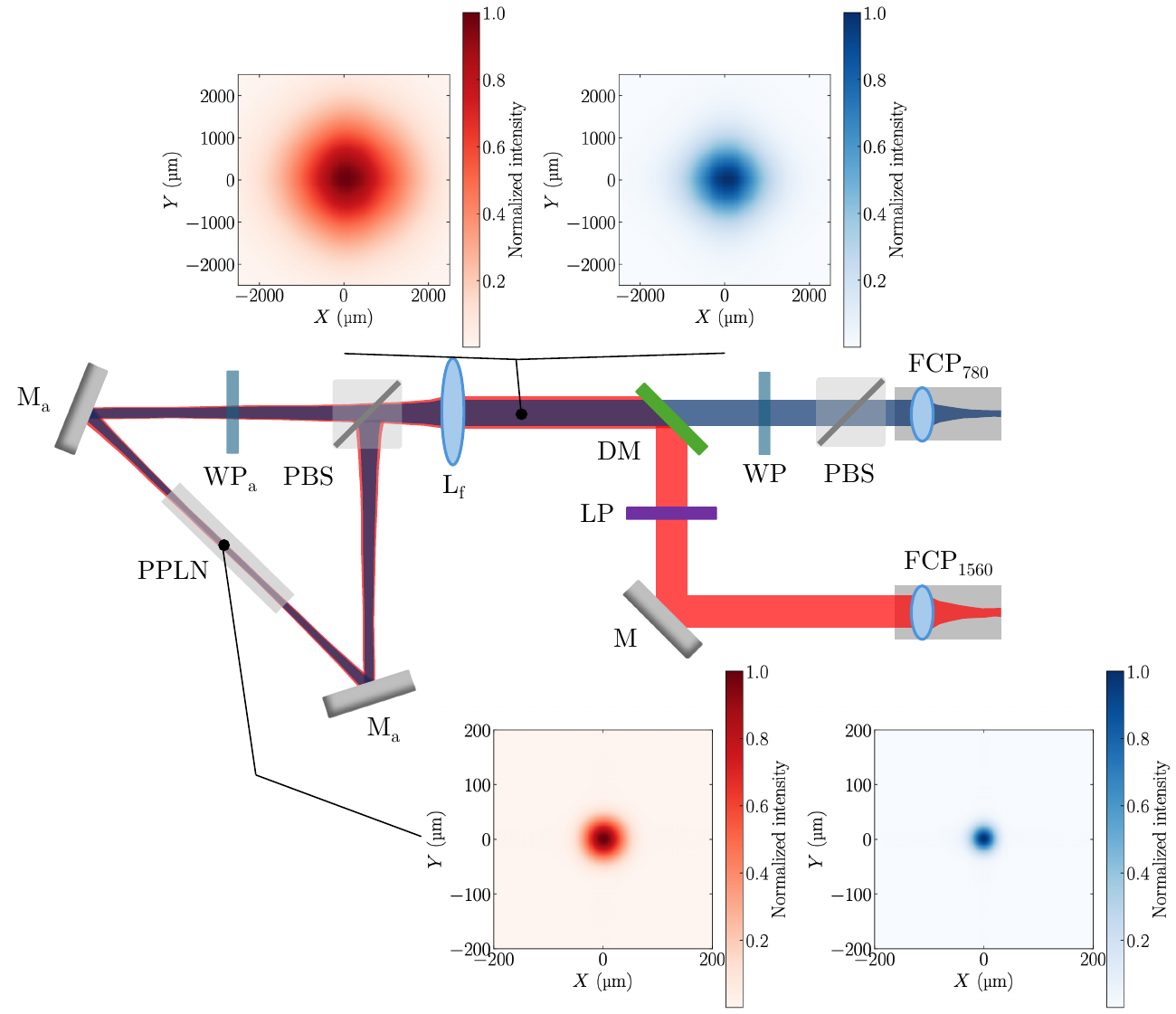}
        (a)
    \end{minipage}
    \hfill
    \begin{minipage}[c]{0.40\textwidth}
        \centering
        \includegraphics[width=\linewidth]{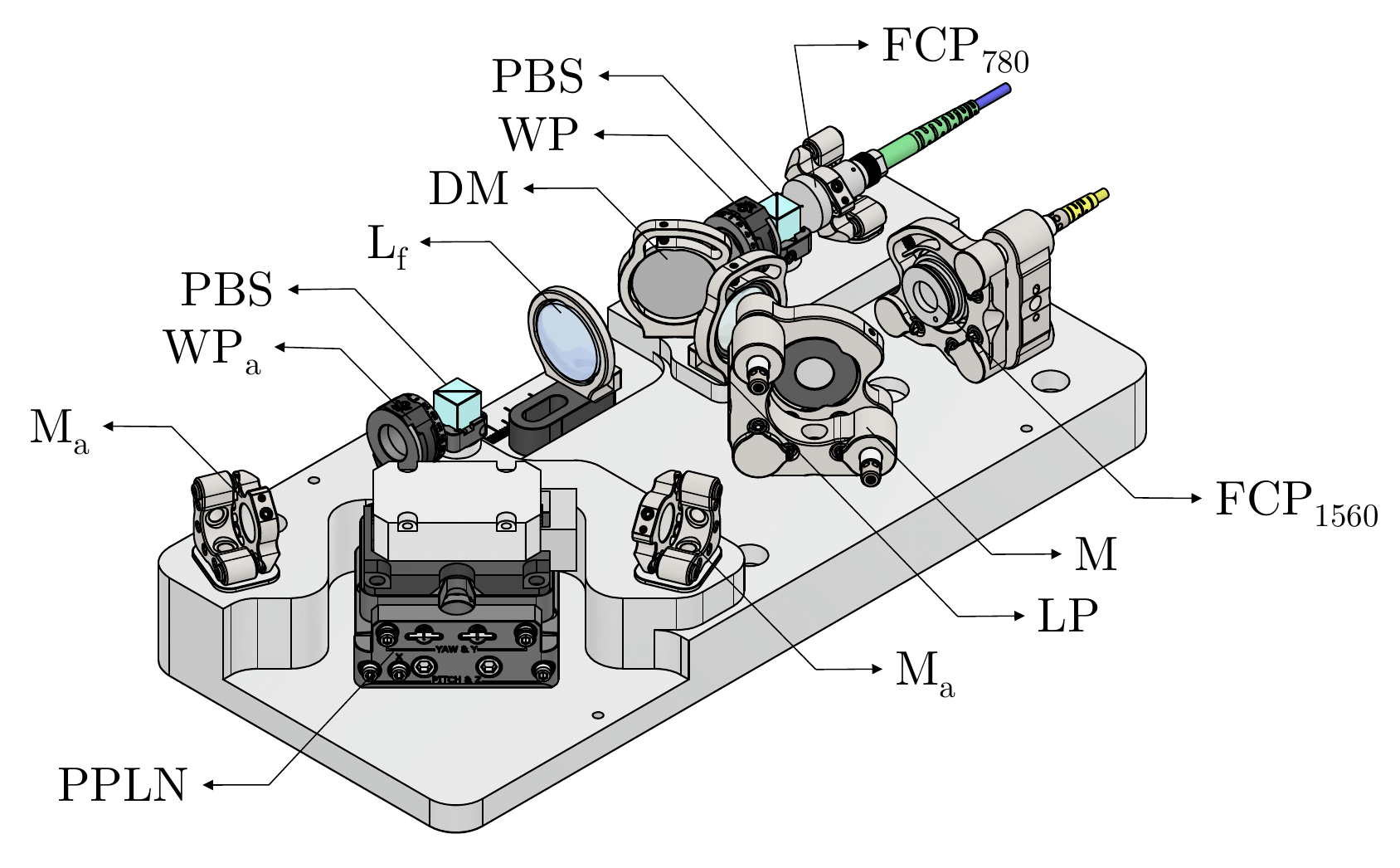}
        (b)
        
        \vspace{0.5em}

        \includegraphics[width=\linewidth]{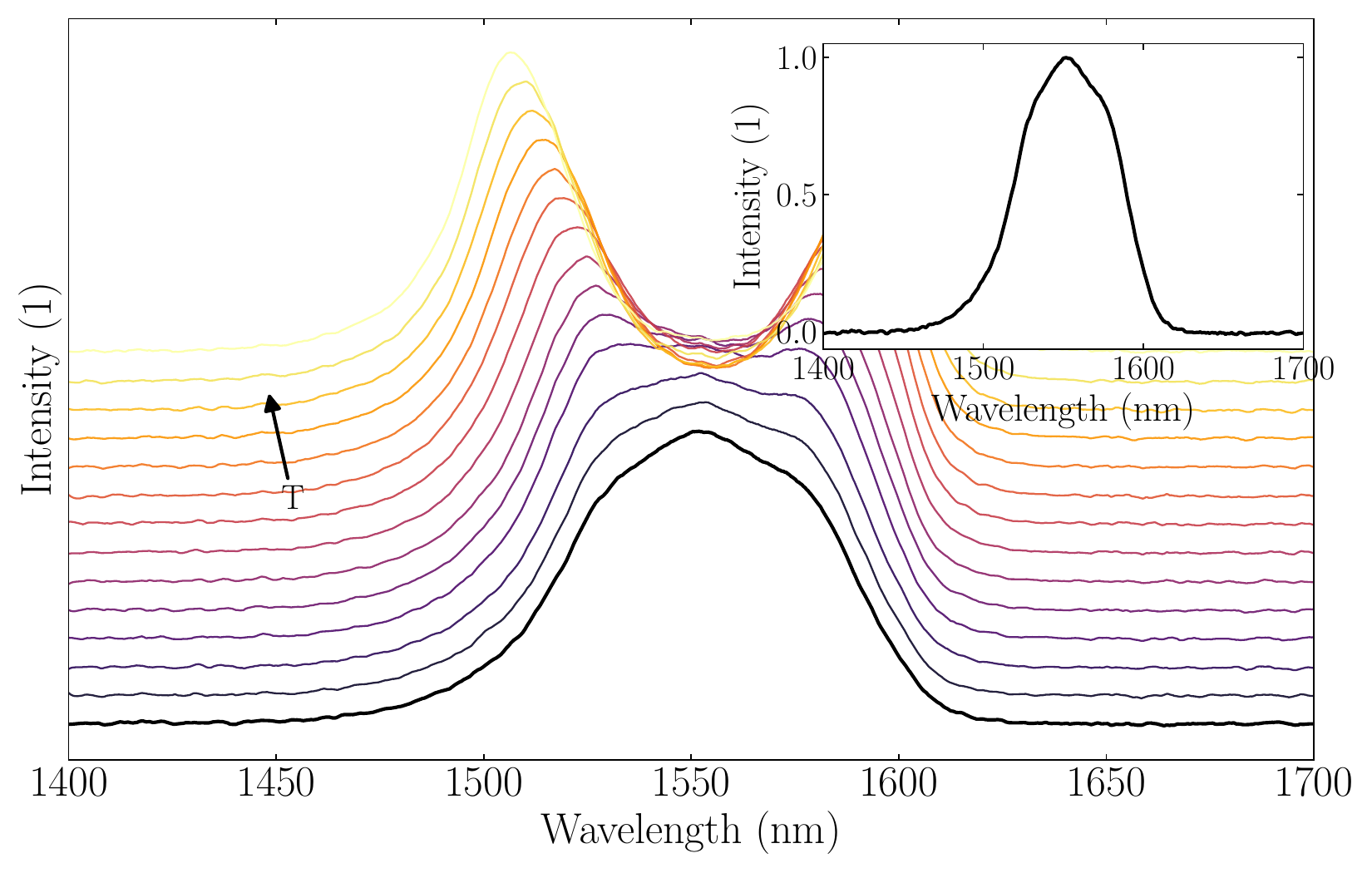}
        (c)
    \end{minipage}

    \caption{\textbf{Robust and high performance polarization entangled photon-pair source based on SPDC in a QPM ppLN crystal} (a) Optical layout for SPDC photon-pair generation. The pump beam (blue) is coupled through the fiber collimation package FCP$_{780}$, with its polarization adjusted using a polarizing beam splitter (PBS) and a wave plate (WP). A dichroic mirror (DM) transmits the pump beam to lens L$_\text{f}$ and a second PBS, which forms the Sagnac interferometer by splitting the pump beam into two counter-propagating paths. The pump is focused onto the center of the temperature-controlled ppLN crystal. The generated SPDC photons (red beam path) are reflected by the DM and collected into FCP$_{1560}$ using mirror M. The inset plots show the beam waists of the pump (blue) and SPDC  (red) beams, respectively. (b) The base plate holds the optical components, reducing the degrees of freedom for easy realignment and shortening the beam paths to make the source intrinsically more stable. (c) Measured SPDC spectra for different ppLN crystal temperatures, showing the temperature-dependent spectral shift. The inset shows the spectrum at the optimum temperature for degenerate photon-pair generation.}
    \label{fig:sagnac}
\end{figure*}
Here, we characterize the main performance metrics, the spectral width, and stability performance by demonstrating successful integration into a deployed eQKD system, marking a step toward practical quantum communication infrastructures. We first introduce the opto-mechanical design of the source, its performance characterization and comparison to state-of-the-art. Building on these results, its suitability for long-distance QKD is investigated via realistic simulations of fiber-based applications.

The SPDC process remains the most established and reliable method for generating entangled photon pairs. The underlying principle is that, through a lowest-order parametric interaction mediated by the second-order nonlinear susceptibility $\chi^{(2)}$ in a nonlinear medium, a strong electromagnetic pump field ($p$) is converted into correlated signal ($s$) and idler ($i$) photon pairs. The conversion efficiency is low \cite{couteau_spontaneous_2018}. As such, practical source designs that generate entangled photon pairs via SPDC must be engineered by careful optimization of mode overlap and optical loss minimization.
The pump laser must be tuned to the quasi-phase-matching condition of the crystal \cite{ortega_spatial_2023}. In addition, its linewidth must be sufficiently narrow, so that SPDC photons of a given wavelength originate from a monochromatic pump. Otherwise, photons of the same wavelength generated from different pump wavelengths would acquire different phases, reducing the phase coherence of the entangled state. Further, the linewidth of the pump constrains the sum of the signal and idler frequencies resulting in strong frequency anti-correlation. To optimize the joint spectral density and thereby maximize the photon-pair generation rate, for an SPDC source with co-propagating Gaussian beams aligned along the principal refractive axes of the nonlinear crystal, the relevant free parameters are the beam waist radii of the interacting fields, denoted by $w_p$, $w_s$, and $w_i$ for the pump, signal, and idler, respectively. The corresponding dimensionless focusing parameters are defined as
\[
\xi_j \equiv \frac{L}{k_j w_j^2}, \qquad j \in {p,s,i},
\]
where $L$ is the length of the nonlinear crystal and $k_j$ is the wave number of field $j$ such that $k_j=n_j\,\omega_j/c$ with refractive index $n_j$ and the optical frequency $\omega_j$. The beam waists were selected to operate close to the focusing conditions that maximize the efficiency of three-wave mixing processes with Gaussian beams, including sum-frequency generation and parametric amplification, as reported in \cite{bennink_optimal_2010, dixon_heralding_2014}. Under these conditions, the optimal focusing parameter is approximately $\xi_{j,\mathrm{opt}} \approx 2.84$.

The optical layout of the source interferometer is shown in Figure~\ref{fig:sagnac}~(a). The base plate holding the optical components is depicted in Figure~\ref{fig:sagnac}~(b). A high-power pump beam is coupled into the setup through a fiber collimation package (FCP) and passes through a polarizing beam splitter (PBS) and a half-wave plate (HWP), defining the polarization state of the pump light. The beam is then transmitted by a dichroic mirror (DM) and focused by a lens L$_\text{f}$ into a temperature-controlled ppLN bulk crystal.
In the Sagnac interferometer, the pump beam is split into two counter-propagating beams of equal intensity. The polarization of one propagation direction is rotated by HWP$_\text{a}$ to align with the nonlinear interaction axis of the ppLN crystal. Under type-0 SPDC, both the signal and idler photons are generated with the same polarization as the pump photon. To generate photon pairs in the polarization-entangled Bell state
\begin{equation}
    \ket{\Phi^+}_{s,i} = \frac{\ket{H}_s\otimes\ket{H}_i + \ket{V}_s\otimes\ket{V}_i}{\sqrt{2}},
    \label{eq:bellphi+}
\end{equation}
the half-wave plate inside the Sagnac loop serves two purposes. The HWP aligns both counter-propagating pump beams with the nonlinear axis of the ppLN crystal for efficient bidirectional SPDC, and rotates the polarization of the generated pairs so that they are orthogonal when they recombine coherently at the PBS.
The pump lens L$_\text{f}$ likewise serves another purpose: It collimates the SPDC photons before they are reflected by DM. An additional mirror (M) helps coupling them into a single-mode output fiber mounted on coupler FCP$_\text{1560}$. The mirror M provides an extra degree of freedom for optimizing fiber-coupling efficiency. 

The main advantage of this design is that it allows a more compact Sagnac loop. A shorter loop is mechanically more stable and permits tighter focusing, since shorter focal lengths can be used. In conventional designs, the different Rayleigh lengths of the pump and the signal/idler modes require different focusing conditions, i.e., different values of $\xi_p$ and $\xi_{s, i}$, and therefore two separate lenses, one for each wavelength. These lenses must act on the spectrally separated beams, so the DM has to sit between the PBS and the lenses. The added distance limits how tightly the beams can be focused. Our design avoids this limitation by using a single lens L$_{\text{f}}$ placed before the DM, which both focuses the pump and collimates the SPDC photons without compromising overall source performance.

The pair generation rate $B$ quantifies the efficiency of the SPDC process. In the ideal case, both photons of a generated pair are collected and detected. Collection efficiency is partially captured by the heralding efficiency, defined as the conditional probability that one photon of a pair is collected and detected given that its partner is also collected and detected. We define $\eta_A$ as the heralding efficiency for photon $A$ being emitted into the collected mode, conditioned on collecting photon $B$, and $\eta_B$ analogously for photon $B$. The combined heralding efficiency is defined as $\eta = \sqrt{\eta_A \eta_B}$.

Experimentally, these metrics are derived from the measured single-count rates $S_A$ and $S_B$ and the two-photon coincidence rate $CC_m$. The detection events corresponding to the single photons are recorded using precise time tagging, and coincidence events are identified by comparing the arrival times of detections. If two detection events occur within a predefined coincidence window $\Delta t_c$, they are classified as a coincidence event. The width of this coincidence window is determined primarily by the timing jitter of the detector system and related electronics. Noise photons can lead to erroneous coincidence detection, so-called accidental coincidences, that are estimated by $CC_a = S_A\,S_B\,\Delta t_c$ and subtracted to yield the net coincidence rate $CC_t = CC_m - CC_a$.
The number of true coincidences is given by the pair generation rate multiplied by the heralding efficiencies of the two photons, $CC_t = B \cdot \eta_A \cdot \eta_B$.
The heralding efficiencies are given by
\begin{align}
    \eta_A &= \frac{CC_t}{S_B}
    &
    \eta_B &= \frac{CC_t}{S_A}
\end{align}
and this allows to compute the pair generation rate as $B = S_A \cdot S_B / CC_t$. It is useful to define a spectral pair generation rate $B_s$ that is equal to $B$ divided by the spectral bandwidth $\Delta\lambda$.

Entangled photon pairs exhibit non-local correlations that can be observed via projective measurements. The polarization contrast visibility quantifies the contrast between the maximum and minimum coincidence rates obtained when analyzing the polarization correlations in a measurement basis. It is defined as
\begin{equation}
    V = \frac{CC_{max}-CC_{min}}{CC_{max}+CC_{min}}
    \label{eq:visibility}
\end{equation}
where $CC_{\max}$ and $CC_{\min}$ denote the maximum and minimum coincidence rates in a given measurement basis, respectively.

The fidelity of a quantum state quantifies the closeness of the experimental quantum state to the state described in (\ref{eq:bellphi+}). From the visibilities in two mutually unbiased bases, e.g. H/V and D/A basis, it can be shown \cite{anwar_entangled_2021} that the lower bound of the Bell-state fidelity is given by
\begin{equation}
    F \geq \frac{1}{2} (V_{HV} + V_{DA}).
\end{equation}

\section{Results}
In the present implementation, a 50\,mm-long QPM-ppLN crystal is pumped at a wavelength of $779.89$\,nm, producing signal and idler photons centered at $1559.79$\,nm. At the degeneracy temperature of the crystal, the source exhibits a broad emission spectrum with a bandwidth of approximately 70\,nm (full-width at half maximum) centered around this wavelength. The measured spectra for different temperatures of the ppLN crystal are shown in Figure~\ref{fig:sagnac}~(c).

The beam waists at the crystal position yield experimentally realized focal parameters of $\xi_p = 2.32$, $\xi_s = 2.25$, and $\xi_i = 2.33$. These values are close to the theoretically optimal focal parameter $\xi_{j,\mathrm{opt}}$. The signal/idler beam waist was determined indirectly. Instead of measuring the SPDC mode itself, we characterized the waist of a beam sent back through the single-mode fiber from FCP$_{1560}$. By reciprocity, this back-propagated beam traces the mode that the fiber collects, so it serves as a proxy for the SPDC mode.

\begin{figure}[htbp]
    \centering
    \includegraphics[width=\linewidth]{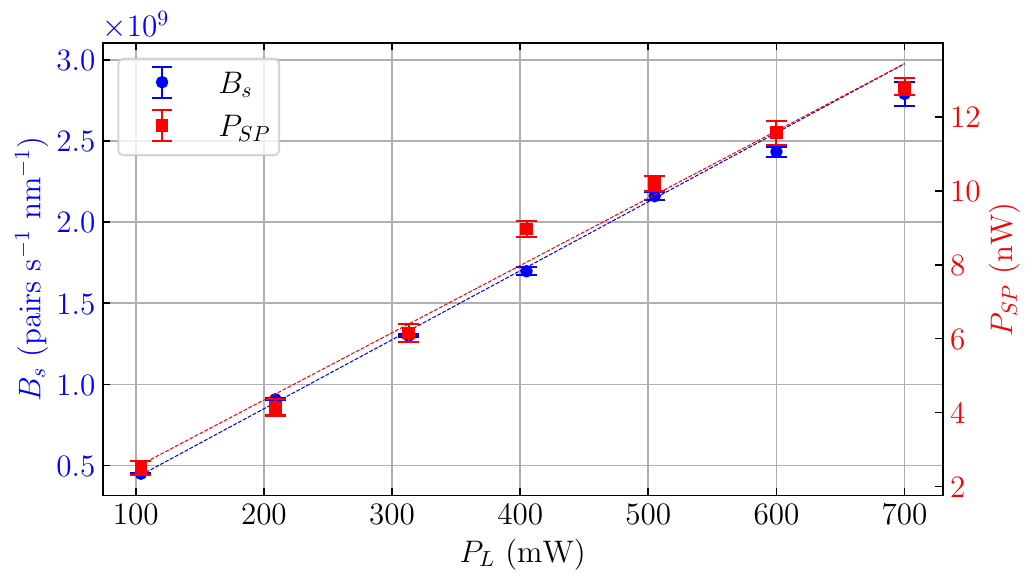}
    \caption{\textbf{Measured pair generation rate $B_s$ and optical power of SPDC signal $P_{SP}$} plotted versus the pump power $P_L$ of up to 700\,mW, demonstrating a spectral pair generation rate of $2.79 \times 10^{9}$\,pairs s$^{-1}$ nm$^{-1}$. The SPDC signal is so bright that it can be measured directly with an off-the-shelf optical power meter; these measurements are denoted as $P_{SP}$.}
    \label{fig:pgr}
\end{figure}

The source achieves a spectral pair generation rate of $B_s=4.24\times 10^{6}$\,pairs s$^{-1}$mW$^{-1}$nm$^{-1}$, measured across pump powers up to $P_p = 700$\,mW with both signal and idler filtered by a 100\,GHz standard dense wavelength-devision multiplexing filter. The intrinsic heralding efficiencies are $\eta_A^i=55.3\%$ and $\eta_B^i=64.8\%$, corrected for detector efficiency and optical losses downstream of the collection fiber in FCP$_{1560}$. At the maximum operating pump power of 700\,mW, the pair generation rate is $B=2.79\times10^{9}$ pairs s$^{-1}$ nm$^{-1}$. The linear dependence of $B$ on the optical pump power, shown in Figure~\ref{fig:pgr}, is consistent with operation in the low-gain regime over the investigated range. 
The high SPDC photon flux can even be measured by means of a conventional off-the-shelf Germanium photodiode optical power sensor. Already at pump powers of 104\,mW, an optical intensity of 2.49\,nW can be measured. At the maximum operating pump powers, the SPDC photon flux coupled into the fiber was measured to be approximately $P_{\mathrm{SPDC}}=12.81$\,nW. This optical power measurement provides an independent indication of the high pair generation rate of the source.

Compared with other interferometric approaches, Sagnac interferometers can achieve particularly high visibilities due to their inherently stable common-path geometry. The visibility quantifies the degree of interference between the counter-propagating optical fields, with the measured visibility in both basis reaching 99.6\,\%. Table~\ref{tab:results} summarizes all relevant metrics.

\begin{table}[htbp]
\caption{Main figures of merit of the reported source of polarization-entangled photon pairs based on type-0 SPDC in a bulk QPM-ppLN.}
\label{tab:results}
\begin{tabular}{@{}>{\raggedright\arraybackslash}p{0.44\columnwidth}>{\raggedright\arraybackslash}p{0.46\columnwidth}@{}}
\hline\hline
Pair generation rate $B$ at $P_L^{\max}$ & \mbox{$2.79\times10^{9}$} pairs\,s$^{-1}$\,nm$^{-1}$ \\
Spectral pair generation rate $B_s$ & \mbox{$4.24\times10^{6}$} pairs\,s$^{-1}$\,mW$^{-1}$\,nm$^{-1}$ \\
Heralding efficiencies $\eta_A$, $\eta_B$ & 0.387, 0.407 \\
Intrinsic heralding efficiency $\eta^i$ & 0.598 \\
Polarization visibility $V$ & 99.6\,\% \\
Fidelity $F$ & $\geq$ 99.5\,\% \\
Emission spectrum at degenerate temperature (FWHM) & $\approx$ 70\,nm \\
Long-term stability (after 9~months) & $V \approx 98.4\,\%$ \\
\hline\hline
\end{tabular}
\end{table}

\input{comparison_inside.tex}


To place our source in context, Figure~\ref{fig:existingwork} compares its performance with state-of-the-art type-0 SPDC entangled-photon-pair sources reported in the literature, plotting the reported heralding efficiency $\eta$ against the spectral pair generation rate $B_s$. The sources are grouped by their operating wavelength regime: both signal and idler in the telecom range \cite{cabrejo-ponce_ghz-pulsed_2022, neumann_experimental_2022, fazili_simple_2024}, one photon in the telecom and one in the visible range \cite{hentschel_three-color_2009, szlachetka_ultrabright_2023, dosan_highly_2025}, or both photons in the visible range \cite{steinlechner_efficient_2014, chen_polarization_2018, lohrmann_broadband_2020, brambila_ultrabright_2023, park_ultrabright_2025}. Bulk ppKTP sources are indicated by circles, bulk ppLN sources by diamonds, and ppLN waveguide implementations by squares. Sources based on waveguides offer a high spectral pair generation rate owing to the strong mode confinement. However, heralding efficiencies are still limited due to fiber coupling losses between waveguide and optical fiber. While some sources achieve exceptionally high spectral pair generation rates or heralding efficiency, our bulk source presents a favorable positions among state-of-the-art implementations, combining competitive performance in both metrics with high fidelity and polarization visibilities. Aligning together with its stable and reproducible operation presented further, these characteristics highlight potential for demanding quantum communication applications. It should be noted that the reported values may vary depending on definitions of filtering the photons and measurement apparatus in each study. For consistency, we use reported values without corrections for optical losses and detector efficiency. Studies that reported spectral pair generation rates per GHz were converted to rates per nm in the telecom C-band.

\subsection{For long-distance eQKD}


To evaluate the suitability of the source for eQKD, its performance was modeled assuming a BBM92 protocol implementation under experimentally realistic operating conditions \cite{neumann_model_2021}. The model to estimate secret key rates incorporates experimentally measured source parameters, including the polarization contrast visibilities, heralding efficiencies, and spectral pair generation rate, together with quantum channel attenuation and the characteristics of superconducting nanowire single-photon detectors (SNSPDs). The detectors are assumed to have a timing jitter of 20\,ps, deadtime of 100\,ns and detector efficiency of 80\,\%. Detector efficiencies and dark count probabilities were assumed to be identical for Alice and Bob. The remaining system parameters were a 100 GHz spectral filtering bandwidth, a detector timing jitter of 20 ps per detector, a dead time of 100 ns, and an error correction efficiency of 1.2 for secret key rate distillation in the asymptotic regime.

\begin{figure}[!htbp]
    \centering
    \includegraphics[width=\linewidth]{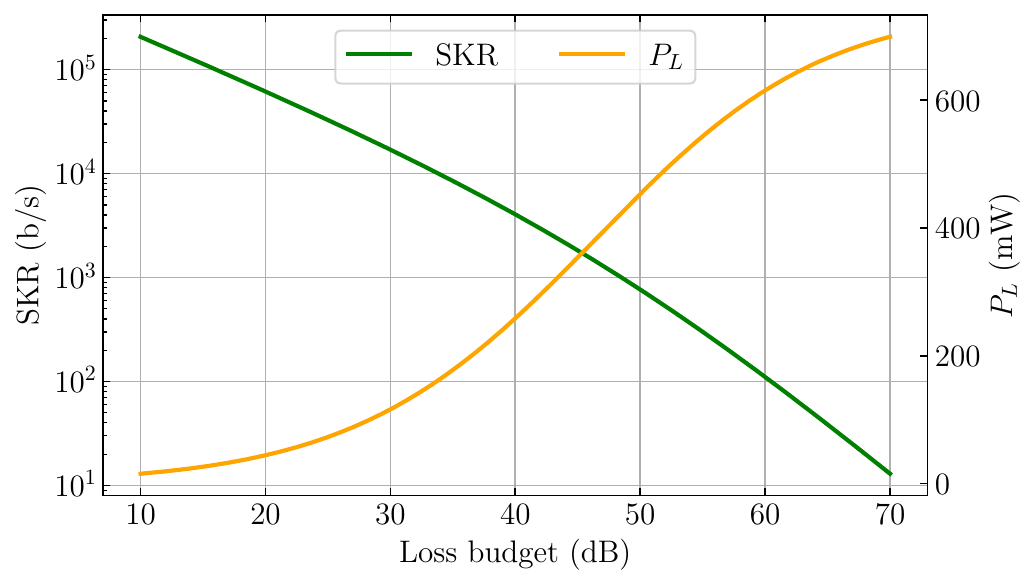}
    \caption{\textbf{Simulation for entanglement-based quantum key distribution} assuming real-world SNSPD detectors and symmetric loss distribution between Alice and Bob. The simulation (based on \cite{neumann_model_2021}) indicates positive secret keys rates (SKR) for a total quantum channel loss budget of up to 70\,dB. The pump power ($P_L$) required for such a long-distance connection stays within the no-damage limits tested in the laboratory.}
    \label{fig:simulation}
\end{figure}

Figure~\ref{fig:simulation} shows the simulated secret key rate (SKR) as a function of total quantum channel loss budget for experimentally relevant source pair generation rates. Simulations indicate that, for the measured source performance, positive SKRs can be achieved even at channel losses approaching 70\,dB. At this loss level, the predicted secret key rate remains on the order of 10\,b/s. Such high-loss scenarios require very high pair rates, since only a small fraction of the generated pairs survives the channel and is detected. The presented source readily provides these rates, reaching $2.79\times10^9$ pairs s$^{-1}$ nm$^{-1}$ at the highest investigated pump power. This suggests its potential for entanglement-based QKD in high-loss scenarios. Also note that the key rate limitation for state-of-the-art eQKD implementations originates from imperfect detection rather than insufficient source brightness, as can be inferred from the saturation trend of the $P_L$ curve; creating more pairs by pumping even harder isn't feasible because the detectors cannot temporally resolve them anymore.

\subsection{Long-term stability}

\begin{figure*}[ht]
    \includegraphics[width=\textwidth]{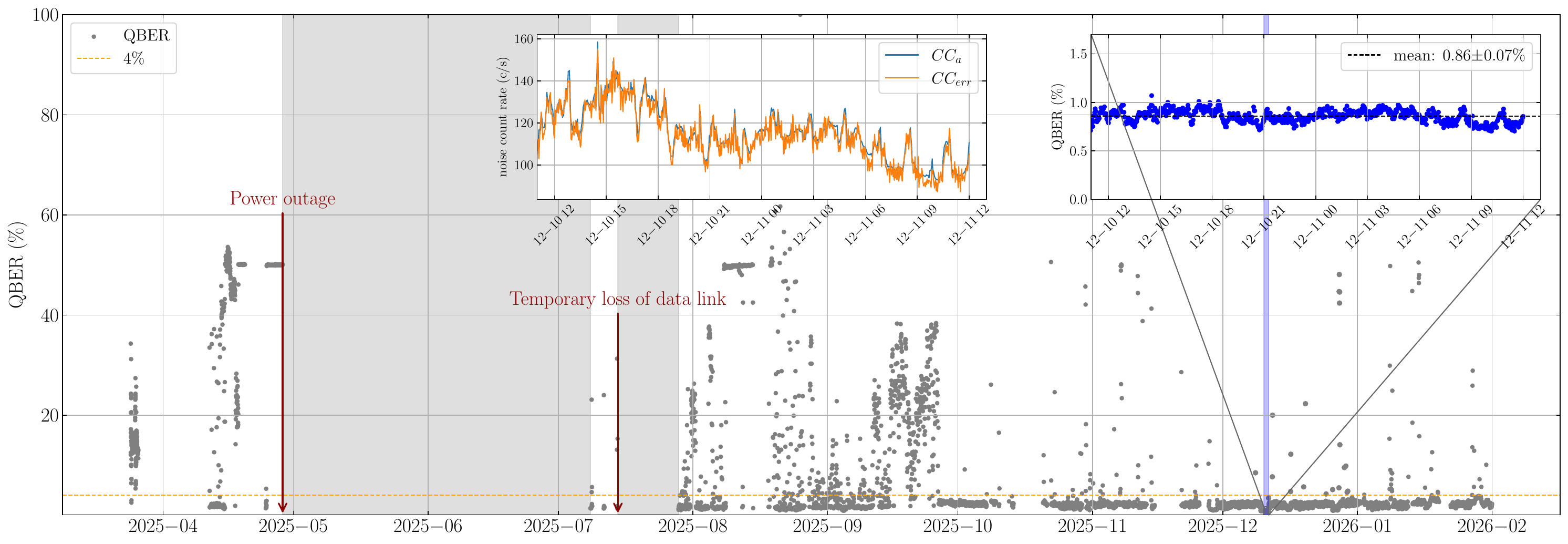}
    \caption{\textbf{Long-term operation of the entanglement-based QKD system over 12 months without optical realignment.} The QBER remained adjustable to below operating threshold of 4\% throughout the deployment and reached $0.86\pm0.07\%$ after approximately nine months of continuous operation. Due to polarization drifts in fiber links an automatic polarization control is necessary to lower QBER such that key creation continues. This is not a source effect. The right inset shows a 24\,h interval around this minimum-QBER period, while the left inset compares the measured number of erroneous coincidence events, given by QBER$ \cdot CC_{m}$, with the estimated accidental coincidence rate calculated from the measured single-count rates and coincidence window. The close agreement indicates that the observed QBER during this period was consistent with accidental coincidences. The interruption in the measurement corresponds to a power outage affecting the deployment site in Madrid.}
    \label{fig:longtermqber}
\end{figure*}

A critical requirement for photon sources intended for practical applications is not only high intrinsic performance but also the ability to maintain that performance over extended periods under realistic operating conditions. 
To assess its long-term stability, we retrieve data from an eQKD application with the source running continuously for 12 months \cite{martin_madqci_2024, sebastian-lombrana_scaling_2026} without requiring realignment. In this configuration, the source was co-located with Alice, while Bob was connected via a deployed optical fiber link, yielding a secure key rate of approximately 12\,kb/s.

Figure~\ref{fig:longtermqber} presents the recorded quantum bit error rate (QBER) over the 12-month deployment. In eQKD, the QBER serves as a key indicator of the overall system performance and is defined as the fraction of erroneous bits in the sifted key. The overall QBER contains three contributions: the source QBER, which reflects the quality of the source alignment; the quantum channel QBER, which arises due to polarization mode dispersion as well as polarization drifts in the fiber; and the accidental-limited QBER, which arises from uncorrelated photons which both lost their partner but arrive within the detector's timing jitter. A polarization controller compensates the quantum channel QBER, and the accidental-limited QBER contribution can be calculated from the temporal correlation's noise floor. The remaining source QBER depends on the source visibility as $QBER_\mathrm{src}=(1-V)/2$. It should be noted that in the eQKD application, active polarization control is employed to compensate for polarization drifts on the quantum channel and maintain the QBER below a predefined operating threshold of 4\,\% rather than to minimize it to the absolute minimum.

To gain further insight, an interval of 24\,h after approximately nine months of continuous operation was analyzed in detail depicted in the inset plot of Figure~\ref{fig:longtermqber}. During this period, the QBER remained as low as 0.84\,\%. This corresponds to a lower bounded $V_{9M}\leq98.28 \pm 0.14\%$ of the source after installation which corresponds to approximately nine months of continuous operation.

The accidental coincidence rate $CC_a$ was estimated from the measured singles count rates and coincidence window, and compared to the number of erroneous coincidence events, $CC_{err} = \mathrm{QBER} \times CC_{tot}$. The close agreement between $CC_a$ and $CC_{err}$ indicates that the observed QBER is accounted for by accidental coincidences, within the experimental uncertainty, leaving no measurable excess error attributable to the source. After the 12-month operation period, the recorded median QBER was 2.13\%, which corresponds to a lower bound of the source visibility of $V_{12M}\leq95.74\%$.

\section{Conclusion}
We have demonstrated a compact and directly fiber-coupled ppLN-based SPDC photon-pair source designed for stable operation in practical quantum communication systems. The optical design was optimized to reduce the number of optical elements and beam-path lengths. This resulted in a single-lens configuration combined with a mechanically engineered base plate without any loss in performance. The source achieves a high pair generation rate combined with high heralding efficiencies. The pair rate scales linearly over a large pump-power range, with no performance reduction even at the highest investigated pump powers, making it suitable for demanding applications. Long-term deployment on the deployed fiber of the MadQCI ecosystem shows that the performance of the source is not limited to laboratory operation but can be maintained under realistic conditions of an eQKD deployment over 12 months of continuous operation. In addition, the long-distance eQKD simulation, assuming state-of-the-art superconducting nanowire single-photon detectors, indicates that the measured source performance is compatible with operation over highly lossy channels, with positive key generation maintained up to approximately 70\,dB of quantum channel loss under realistic assumptions.

The asymmetric heralding efficiencies need to be  investigated further. Adaptation of the focusing conditions and fiber coupling could reduce the asymmetry. Other contributions may arise from differences in the optical components. The dominant limitation is expected to arise from coupling into the single-mode fiber. In particular, measured beam profiles showed a slight truncation of the beam, suggesting that the spatial mode overlap with the fiber mode is not yet fully optimized. Although the pair generation rate $B$ reaches $2.79\times10^9$ pairs s$^{-1}$ nm$^{-1}$ at the highest investigated pump power, multipair emission is not expected to be a limiting factor under CW pumping. In this regime, the pair emission is spread over many temporal modes, as the photon coherence time is much shorter than the average time between pairs within the collected bandwidth. Accidental coincidences originating from imperfect temporal detection are therefore the dominant background, not true multipair events.

While the underlying SPDC process and ppLN-based photon-pair generation are well established, this work demonstrates how careful optical and mechanical design can combine high source performance with the stability required for long-term operation in deployed quantum communication systems. Looking ahead, the broad emission spectrum of the presented source offers several avenues for further development. By narrowing the spectral filtering from the demonstrated 100\,GHz bandwidth to 50\,GHz, the source could be adapted to applications in long-distance quantum key distribution, where narrower spectral modes can improve the achievable visibility. Furthermore, if free-space separation of the signal and idler photons is desired, the current scheme could be readily adjusted by changing to a non-degenerate spectrum and replacing the mirror M by an appropriate dichroic mirror, enabling the two photons to be separated directly in free space.
The fiber-coupled pump laser architecture provides a straightforward route towards implementing a pulsed pump laser. Replacing the current continuous-wave laser with a suitable fiber-coupled pulsed source would provide a well-defined temporal reference while retaining the existing fiber-coupled interface. This could enable the synchronization of multiple SPDC sources and, in particular, provide a route towards combining two such sources for the generation of four-photon states.

\section*{Acknowledgments}
M.S. is a doctoral researcher at the University of Vienna, Austria and carried out this work at zerothird GmbH. We thank Joel Muela for providing Figure~\ref{fig:sagnac}~(b).

\section*{Author Contribution}
M.S. conceived the study, co-designed and constructed the source, performed the characterization, and wrote the manuscript. M.S. and J.B. carried out the spectral measurements, and J.B. and S.P.N. contributed to the writing. M.H. designed the mechanical components together with M.S., S.P.N., and T.H.; T.H. and S.P.N. also co-designed the source. J.J.R. supported the long-term data acquisition. R.U. and S.P.N. supervised the work, and P.W. provided academic supervision for M.S. and J.B and the laboratory facilities for the spectral measurements.

\section*{Data availability}
The data supporting the findings of this study are available upon reasonable request.

\section*{Competing interests}
M.S., T.H., and S.N. are employees of zerothird GmbH, which develops and commercializes entanglement-based quantum key distribution systems. M.H. and R.U. are affiliated with qtlabs GmbH. J.B. J.J.M, and P.W. declare no competing interests.


\bibliographystyle{naturemag}
\bibliography{sourceLib}

\end{document}

%% file: comparison_inside.tex
\begin{figure}[!htbp]
  \centering
  \begingroup
  \definecolor{cmporange}{HTML}{D55E00}
  \definecolor{cmpblue}{HTML}{0072B2}
  \definecolor{cmpgreen}{HTML}{009E73}
  \fontsize{7pt}{9.6pt}\selectfont
  \begin{overpic}[width=\linewidth,percent]{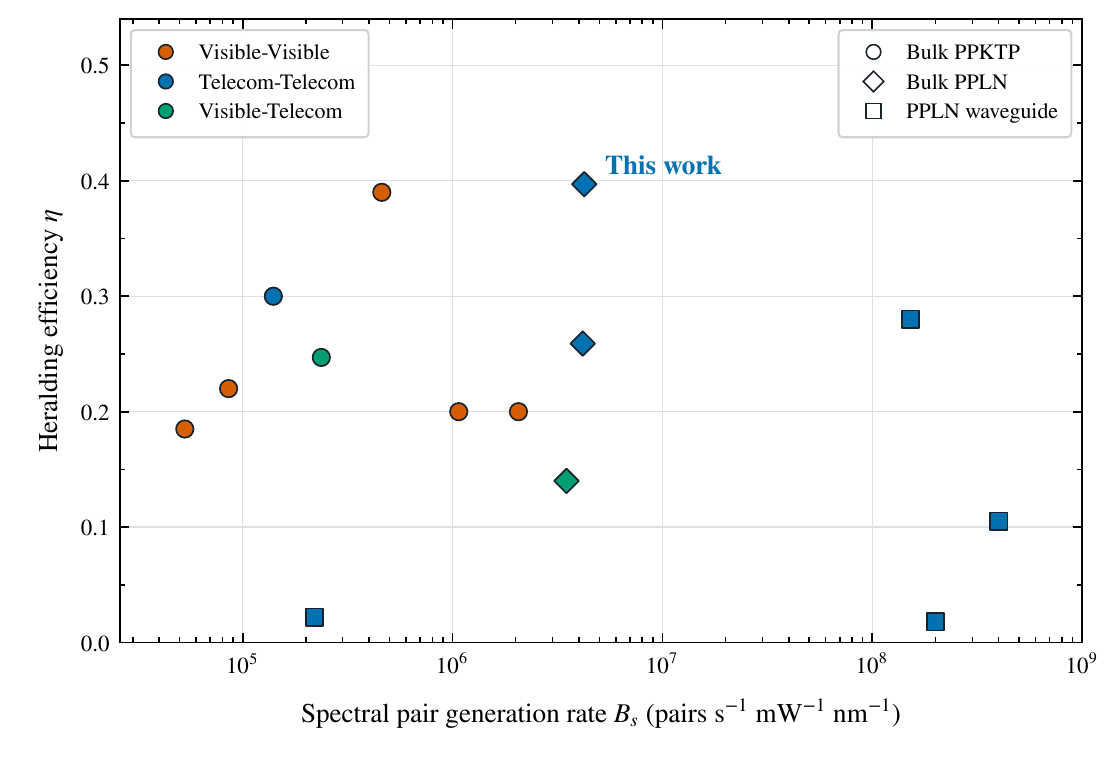}
    \put(35.70864,51.59347){\makebox(0,0)[l]{\color{cmporange}\hypersetup{citecolor=cmporange}\cite{steinlechner_efficient_2014}}}
    \put(19.14722,26.97008){\makebox(0,0)[r]{\color{cmporange}\hypersetup{citecolor=cmporange}\cite{chen_polarization_2018}}}
    \put(19.09507,34.66141){\makebox(0,0)[r]{\color{cmporange}\hypersetup{citecolor=cmporange}\cite{lohrmann_broadband_2020}}}
    \put(53.99976,37.96730){\makebox(0,0)[l]{\color{cmpblue}\hypersetup{citecolor=cmpblue}\cite{neumann_experimental_2022}}}
    \put(29.83899,13.52775){\makebox(0,0)[l]{\color{cmpblue}\hypersetup{citecolor=cmpblue}\cite{cabrejo-ponce_ghz-pulsed_2022}}}
    \put(39.80037,33.91066){\makebox(0,0)[r]{\color{cmporange}\hypersetup{citecolor=cmporange}\cite{brambila_ultrabright_2023}}}
    \put(23.12418,42.23198){\makebox(0,0)[r]{\color{cmpblue}\hypersetup{citecolor=cmpblue}\cite{fazili_simple_2024}}}
    \put(48.20840,31.83033){\makebox(0,0)[l]{\color{cmporange}\hypersetup{citecolor=cmporange}\cite{park_ultrabright_2025}}}
    \put(30.45025,46.39264){\makebox(0,0)[l]{\color{cmpgreen}\hypersetup{citecolor=cmpgreen}\cite{dosan_highly_2025}}}
    \put(52.53786,25.58934){\makebox(0,0)[l]{\color{cmpgreen}\hypersetup{citecolor=cmpgreen}\cite{szlachetka_ultrabright_2023}}}
    \put(88.47370,21.94876){\makebox(0,0)[r]{\color{cmpblue}\hypersetup{citecolor=cmpblue}\cite{zhuang_ultrabright_2025}}}
    \put(83.53067,40.15165){\makebox(0,0)[l]{\color{cmpblue}\hypersetup{citecolor=cmpblue}\cite{troisi_high-brightness_2026}}}
    \put(85.78441,13.46239){\makebox(0,0)[l]{\color{cmpblue}\hypersetup{citecolor=cmpblue}\cite{yadav_high-brightness_2024}}}
  \end{overpic}
  \endgroup
  \caption{\textbf{Performance comparison to recent existing work.} The heralding efficiency $\eta = \sqrt{\eta_A\eta_B}$ is plotted against the spectral pair generation rate $B_s$ of state-of-the-art type-0 SPDC entangled photon-pair sources. Colours distinguish wavelength groups and marker shapes identify the SPDC platform.}
  \label{fig:existingwork}
\end{figure}